\documentclass[superscriptaddress,prd,showpacs,aps,12pt]{revtex4}
\usepackage{txfonts}
\usepackage{wasysym}
\usepackage{graphicx,epsfig,amssymb}
\usepackage{multirow}
\usepackage{verbatim}

\newcommand{\newc}{\newcommand}	
\newc{\gsim}{\lower.7ex\hbox{$\;\stackrel{\textstyle>}{\sim}\;$}}
\newc{\lsim}{\lower.7ex\hbox{$\;\stackrel{\textstyle<}{\sim}\;$}}

\def\beq{\begin{equation}}
\def\eeq{\end{equation}}
\def\beqn{\begin{eqnarray}}
\def\eeqn{\end{eqnarray}}

\begin{document}
\title{Search for $Z'$ via $ZH$ associated production at LHC}

\author{Hong-Lei Li}\email{sps\_lihl@ujn.edu.cn}\affiliation{School of Physics and Technology, University of Jinan, Jinan Shandong 250022,  China}
\affiliation{Department of Physics, University of the Chinese Academy of Sciences, Beijing 100049,  China}

\author{Zong-Guo Si}\affiliation{School of Physics, Shandong University, Jinan Shandong 250100,  China}
\author{Xiu-Yi Yang}\affiliation{School of Physics, Shandong University, Jinan Shandong 250100,  China}
\affiliation{School of Science, University of Science and Technology Liaoning, Anshan, Liaoning 114051, China}
\author{Zhong-Juan Yang}\affiliation{School of Physics and Chemistry, Henan Polytechnic University, Jiaozuo Henan, 454003, China}
\author{Ya-Juan Zheng}\affiliation{CASTS, CTS and Department of Physics, National Taiwan University, Taipei 10617, China}
\begin{abstract}
Many new physics models predict the existence of extra neutral gauge bosons ($Z'$). Inspired by the recent development on Higgs search, we study the properties of $Z'$ via the Higgs and $Z$ boson associated production. The couplings of $Z'$ to quarks can be investigated through $pp \to Z'\to ZH\to  l^+l^-b \bar b$ process, which also provides extraordinary signal for understanding the properties of $Z'ZH$ interaction. The standard model background processes can be significantly suppressed by adopting appropriate kinematic cuts. The charged lepton angular distributions are related to the ratio of chiral couplings of $Z'$ to quarks via $Z'ZH$ interaction.

\end{abstract}
\maketitle	
\section{Introduction}
The ATLAS and CMS collaborations at LHC have reported the discovery of a  SM-like Higgs boson with mass around 125 GeV recently ~\cite{ATLAS-Higgs,CMS-Higgs}. It is timing to study the physics related to the Higgs boson within and beyond SM. In some of the simplest extensions of the SM, such as the simple extra $U(1)'$ gauge symmetry~\cite{Robinett:1981yz,Hewett:1988xc,London:1986dk}, the left-right models ~\cite{Pati:1974yy,Mohapatra:1974hk,Mohapatra:1974gc,Senjanovic:1975rk,Mohapatra:1977mj} and even the string theory~\cite{Cvetic:1995rj,Cvetic:1996mf}, etc., a new neutral gauge boson $Z'$ is introduced. The crucial test for these models is to search for the $Z'$ production signal and to study its related properties.

$Z'$ boson is widely investigated at both LEP and hadron colliders. Constraints on $Z'$ mass and coupling strength with the SM particles have been obtained through $e^+e^- \to f \bar f$ process at the electron-positron collider~\cite{Chiappetta:1996km,Barger:1996kr,Lynch:2000md}. Especially, the precision measurements at the $Z$ pole give limit on the $Z-Z'$ mixing~\cite{mixing}. Bounds on several models containing extra neutral gauge bosons have been set by both the CDF and D0 experiments by measuring high energy lepton pair production cross sections at Tevatron~\cite{Abazov:2010ti, Aaltonen:2008vx}.  Due to the high luminosity and collision energy, the large hadron collider (LHC) implies more promising potential to observe heavy gauge bosons. More recently, the $e^+e^-$, $\mu^+\mu^-$, $\tau^+\tau^-$ final states, along with the $Z'$ decay into jets which suffers from large QCD backgrounds, have also been studied at LHC~\cite{Aad:2012cg}. $Z'$ boson in association with vector bosons as well as top quark production is explored at LHC, and is important for disentangling the origin of electroweak symmetry breaking. 

For  theoretical reviews of $Z'$ boson physics, we refer to ~\cite{Langacker:2008yv,Rizzo:2006nw,Leike:1998wr}. Based on specific new physics models, $Z'$ phenomenological signatures have been studied in the literatures ~\cite{Agashe:2009bb,Agashe:2007ki,Carena:2004xs,Barger:2003hg,Feldman:2007wj,Hebecker:2001wq}. In the extensively studied Drell-Yan process, clear signatures for $Z'$ boson can be reconstructed, while it is no picnic investigating the properties of $Z'$ coupling to quarks because of the inevitable mixing from its coupling to leptons. Another important process is $pp\to Z'\to ZH$, which can offer an opportunity to study  the properties of $Z'q\bar q$ as well as $Z'ZH$ interaction.

 In this paper, we investigate the $Z'$ signal via the $ZH$ associated production at LHC with subsequent decay of $Z\to l^+l^-$ and $H\to b\bar{b}$, which will shed light on the understanding of $Z'q{\bar q}$ and $Z'ZH$ interactions. It shows that the couplings of  $Z'$ boson to quarks  are related to the charged leptons angular distributions through $Z'ZH$ interaction. 

This paper is organized as follows. In Sec.II, experimental constraints and theoretical framework are briefly introduced. Sec.III is devoted to the numerical analysis of $Z'$ mediated $ZH$ production at LHC and the corresponding SM backgrounds are considered as well. Finally, a short summary is given.
 
\section{Experimental constraints and Theoretical Framework}

In this part we collect the constraints on the $Z'$ mass from $pp$ and $p\bar p$ colliders. The CMS collaboration excludes leptophobic $Z'$ resonances of masses $m_{Z'}<1.3$ TeV for a width $\Gamma_{Z'}=0.012m_{Z'}$ in a search for heavy resonances decaying into $t\bar t$ pairs with subsequent leptonic decay of both top quark and antiquark process~\cite{2012rq}. The ATLAS collaboration investigates a massive resonance decaying into $t \bar t$ pairs in the fully hadronic final state, and excludes the leptophobic $Z'$ boson model with masses $0.70 < m_{Z'} < 1.00$ TeV and $1.28 < m_{Z'} < 1.32$ TeV as well~\cite{:2012qa}. Based on the analysis of $pp$ collisions at a center-of-mass energy of 8 TeV corresponding to an integrated luminosity of approximately 5.9 ($e^+e^-$) / 6.1 ($\mu^+\mu^-$) $fb^{-1}$, a sequential SM $Z'$ boson is excluded at 95\% C.L. for masses below 2.39 TeV in the electron channel ($e^+e^-$), 2.19 TeV in the muon channel ($\mu^+\mu^-$), and 2.49 TeV in the two channels 
combined ($l^+l^-$)~\cite{a-dilepton}. 
Both D0 and CDF collaborations search for a heavy neutral gauge boson in the $e^+e^-$ channel  of $p\bar p$ collisions at $\sqrt{s}= 1.96$ TeV. A lower mass limit of 1.023 TeV and 0.963 TeV for the sequential SM $Z'$ boson  is presented respectively~\cite{Abazov:2010ti,Aaltonen:2008vx}.
Using constraints from the precision electroweak (EW) data, the lower mass limit on extra neutral boson $Z'_{LR}$ in left-right symmetric models is around 1 TeV~\cite{Erler:2009jh}. The results for various $E_6$-motivated $Z'$ boson are reported from ATLAS  with a lower mass limits of $1.49-1.64$ TeV~\cite{Collaboration:2011dca}, and mass less than $2.09-2.24$ TeV  $Z'$ bosons are excluded by CMS. According to the CMS analysis, a $Z'$ with SM-like couplings can be excluded below 2.59 TeV and the superstring-inspired $Z'_{\psi}$ below 2.26 TeV~\cite{c-dilepton}. However, the $Z'$ mass constraints discussed above depend on the free parameters such as its decay width and branching ratios, and still can be loosened to some extent. 

Considering the existence of $Z'$, the weak neutral current interactions related to the fermions are described by the Lagrangian
\begin{eqnarray}
-{\cal{L}}_{NC}&=& \left[ \bar q\gamma^\mu(g_L^{q}P_L+g_R^{q}P_R)q + \bar l \gamma^\mu(g_L^lP_L+g_R^lP_R)l \right] Z_{\mu}  \nonumber \\
&+& \left[ \bar q \gamma^\mu(g_L^{\prime q}P_L+g_R^{\prime q}P_R)q +\bar l \gamma^\mu(g_L^{\prime l}P_L+g_R^{\prime l}P_R)l \right] Z'_{\mu}.
\label{ffzcoupling}
\end{eqnarray}
where $P_{L,R}=(1\mp \gamma_5) /\ 2$ are the left and right chiral projections and $g_{L,R}^i$ ($g_{L,R}^{\prime i}$) is the chiral couplings of $Z$ ($Z'$) boson to corresponding fermions. $g_{L,R}^i=g/\cos\theta_W(T^i_{3L,R}-\sin^2\theta_W Q^i)$ and $T_3$ ($Q$) is the third component of weak isospin  (charge) of the corresponding particle.  We neglect the $Z-Z'$ mixing in the following, and refer to e.g.~\cite{he,babu,jusak,ilc,Beringer:1900zz} and references therein for corresponding discussions. From Eq.~(\ref{ffzcoupling}), one can obtain the $Z'$ decay width into fermion
\begin{eqnarray}
&\Gamma_{Z'\to f \bar f}=&\frac{C_fm_{Z'}}{24\pi} \left( g_L^{\prime f 2}+g_R^{\prime f2}\right ), \\
&\Gamma_{Z'\to t \bar t}=&\frac{m_{Z'}}{8\pi}\left(1-\frac{m_t^2}{m_{Z'}^2}\right )\sqrt{1-\frac{4m_t^2}{m_{Z'}^2}}\left(g_L^{\prime t 2}+g_R^{\prime t2}\right ),
\end{eqnarray} 
where $C_f$ is the color factor (1 for leptons and 3 for quarks), $m_t$ is the mass of top quark and the light fermion masses have been neglected.

The  $Z'ZH$ interaction can be extracted from 
\begin{eqnarray}
-{\cal{L}}_{kin}
&\equiv&g_{ZZH}Z_{\mu}Z^{ \mu}H+ g_{Z'ZH}Z'_{\mu}Z^{\mu}H+  g_{Z'Z'H}Z'_{ \mu}Z^{\prime \mu}H + \cdots ,
\end{eqnarray}
where $g_{VVH}$ stands for the coupling strength of $VVH$ interaction. We set $g_{Z'ZH}=g_{ZZH}/2$ and $g_{ZZH}=gm_Z/\cos\theta_W$ in the following discussion. 
Thus for a  $Z'$ boson with $m_{Z'}>m_{H}+m_Z$, the partial decay width is
\begin{equation}
\Gamma_{Z'\to ZH}= \frac {g_{Z'ZH}^2}{48\pi m_{Z'}^3}[2+\frac {(m_{Z'}^2+m_Z^2-m_H^2)^2}{4m_{Z'}^2m_Z^2}]\sqrt{[m_{Z'}^2-(m_Z+m_H)^2][m_{Z'}^2-(m_Z-m_H)^2]}.
\end{equation}

\begin{figure}
\begin{centering}
\includegraphics[clip,width=8cm,height=8cm,keepaspectratio]{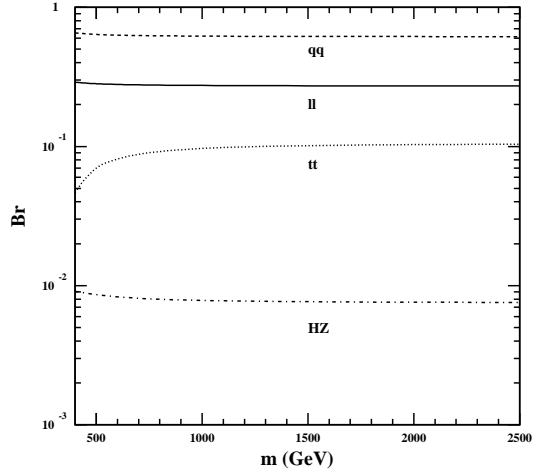}
\caption{$Z'$ decay branching ratio versus $m_{Z'}$.}
\label{fig:br}
\par\end{centering}
\end{figure}

We show the $Z'$ boson decay branching ratio with various mass values in Fig.~\ref{fig:br}.  In this paper, the couplings of $Z'$ to fermions are set to be the same as those of Z boson without special declaration.
 The dominant decay channel is $Z'\to f \bar f$ modes. The branching ratio of $l^+l^-$ channel is about thirty percent, which can be a good channel for discovery of $Z'$ boson at LHC due to the outstanding detector performance, and $ZH$ channel branching ratio is about one percent.  

\begin{figure}
\begin{centering}
\begin{tabular}{c}
\includegraphics[clip,width=8cm,height=8cm,keepaspectratio]{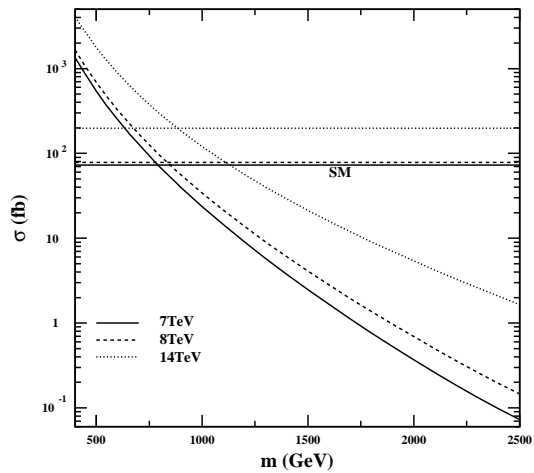}  \tabularnewline
\end{tabular}
\caption{The cross section for the process $pp\to Z' \to ZH$ with $m_{Z'}$ at (a) 7 TeV, (b) 8 TeV and (c) 14 TeV LHC. The straight lines stand for the contribution from $pp\to Z \to ZH$.}
\label{fig:cs-ZH} 
\par\end{centering}
\begin{centering}
\par\end{centering}
\centering{} 
\end{figure}

However at LHC Higgs in associated with $Z$ boson production plays an important role and $q \bar q \to Z' \to ZH$ process contributes to it. We display the cross section of $ZH$ associated production at LHC in Fig.~\ref{fig:cs-ZH}. The cross section of $ZH$ production from $Z'$ boson is above 100 (1) $fb$ for $m_{Z'}=1$ (1.5) TeV at LHC with $\sqrt{s}=14$ TeV, which leads to considerable effects compare with that from $Z$ boson, so it is possible to study $ZH$ associated process via $Z'$ production at LHC.

With the  $Z$ boson leptonic decay modes,
the corresponding matrix element square for the process
\begin{equation}
 q(p_1)\bar q(p_2)\rightarrow Z/Z'\rightarrow ZH \rightarrow l^+(p_3)l^-(p_4)H
\end{equation} 
is given by
\begin{eqnarray}
|\mathcal{M}|^2 &=& \frac{16}{N_c N_s\mathcal{P}_{2Z} } 
 \left \{ \frac{g_{Z'ZH}^2}{\mathcal{P}_{1Z'}}\left [ ({g_L^{l}}^2 {g_L^{\prime q}}^2+{g_R^l}^2 {g_R^{\prime q}}^2)\mathcal{A}+ 
({g_R^l}^2 {g_L^{\prime q}}^2+{g_L^l}^2 {g_R^{\prime q}}^2)\mathcal{B}\right ] \right. \nonumber  \\
&+&\frac{g_{ZZH}^2}{\mathcal{P}_{1Z}}\left [ ({g_L^{l}}^2 {g_L^{ q}}^2+{g_R^l}^2 {g_R^{ q}}^2)\mathcal{A}+ ({g_R^l}^2 {g_L^{ q}}^2+{g_L^l}^2 {g_R^{ q}}^2)\mathcal{B}\right ] \nonumber  \\
 &+&\left. \frac{g_{ZZH}g_{Z'ZH}\mathcal{T}}{\mathcal{P}_{1Z}\mathcal{P}_{1Z'}}
 \left [ ({g_L^{l}}^2 g_L^q g_L^{\prime q}+{g_R^l}^2 g_R^q g_R^{\prime q})\mathcal{A}+ ({g_R^l}^2 g_L^q g_L^{\prime q}+{g_L^l}^2 g_R^q g_R^{\prime q})\mathcal{B}\right ]  \right \}, 
 \label{msquare}
\end{eqnarray}
with $\mathcal{T}=2(s_1-m_Z^2)(s_1-m_{Z'}^2)+\Gamma_{Z}m_{Z}\Gamma_{Z'}m_{Z'}, ~\mathcal{A}=(p_1\cdot p_3)(p_2\cdot p_4), ~\mathcal{B}=(p_1\cdot p_4)(p_2\cdot p_3), ~\mathcal{P}_{ij}=(s_i-m_j^2)^2+\Gamma_j^2m_j^2, ~i=1,~2 $ and $j=Z,~ Z'$,
where $s_1=(p_1+p_2)^2$ is the energy square in center-of-mass system, $s_2=(p_3+ p_4)^2$ is the invariant mass square of the final leptons, $N_c =3$ and $N_s=4$ are the color and spin factors respectively. From the expression of Eq.~(\ref{msquare}), one can notice that the couplings of $Z'$ to quarks are related to the angle between the momentum of lepton and the initial quark.  

Considering the Higgs hadronic decay, one can find that the partonic level final state is $l^+ l^-b\bar b$. Following the analytical method in our previous works on new boson production~\cite{Han:2012vk,Gong:2012ru,Bao:2011sy,Bao:2011nh,Gopalakrishna:2010xm,Agashe:2007ki,Agashe:2009bb}, we focus on the property of the final lepton angular distribution. The  total momentum of the final particles system is defined as ${\bf p}={\bf p_b}+{\bf p_{\bar b}}+{\bf p_{l^+}}+{\bf p_{l^-}}$ in the laboratory frame. The angle between the 3-momentum ${\bf p}^*_{l^-}$ of the lepton in the $Z$ boson rest frame and  ${\bf p}$ is
\begin{equation}
\cos\theta =\frac {{\bf p}^*_{l^-}\cdot{\bf p}}{|{\bf p}^*_{l^-}|\cdot|{\bf p}|}.
\end{equation} 

\begin{figure}
\begin{centering}
\begin{tabular}{c}
\includegraphics[clip,width=14cm,height=10cm,keepaspectratio]{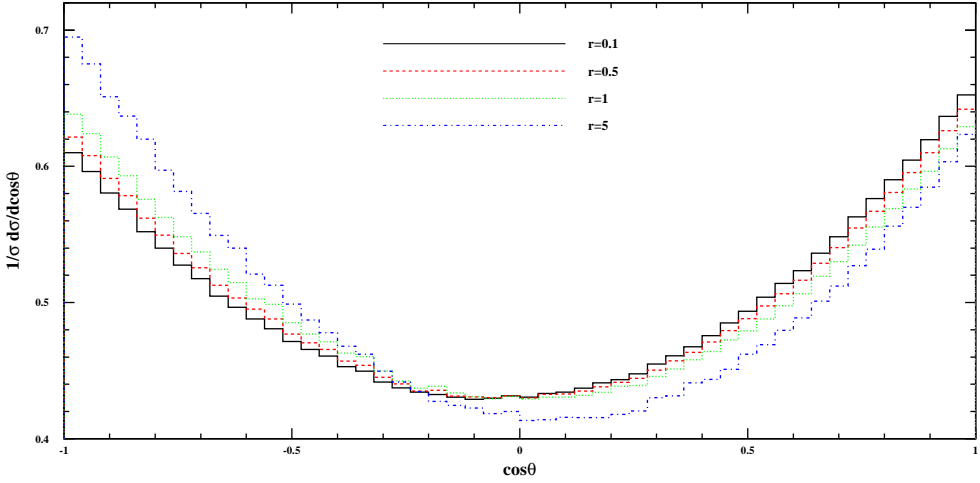}  \tabularnewline
\end{tabular}
\caption{The angular distribution for charged lepton ($l^-$). The mass of $Z'$ is set at 1.5 TeV.}
\label{fig:theta-r}
\end{centering}
\end{figure}

We show distributions of $1/\sigma d\sigma/d\cos \theta$ versus $\cos \theta$ with different $r$ value in Fig.~\ref{fig:theta-r}, where $r$ is defined as $r=g_L^{\prime q}/g_R^{\prime q} $. To give a simplified picture, we set $g_L^{\prime u}=g_L^u$ and $g_L^{\prime u}/g_L^{\prime d}=g_L^{u}/g_L^{d}$. One can find that the charged leptons have large probability to move against the $Z'$ boson boosting direction with large $r$. Therefore, from this kind of angular distribution, one can extract useful information for understanding $Z'q\bar q$ interaction.

%
\section{Collider analysis}
%
$Z'$ boson searches at LHC have been performed by ATLAS and CMS collaborations in the dilepton final states. $ZH$ production via $Z$ boson also elaborately investigated for the search of Higgs boson, thus $Z'$ mediated production
\begin{equation}
 pp\rightarrow Z'\rightarrow ZH \rightarrow l^+l^-b\bar b
\label{eq:pp2llbb}
\end{equation}
will be an important process for Higgs searches beyond the SM.
\begin{figure}
\begin{center}
\begin{tabular}{ccc}
\includegraphics[clip,width=5cm,height=6cm,keepaspectratio]{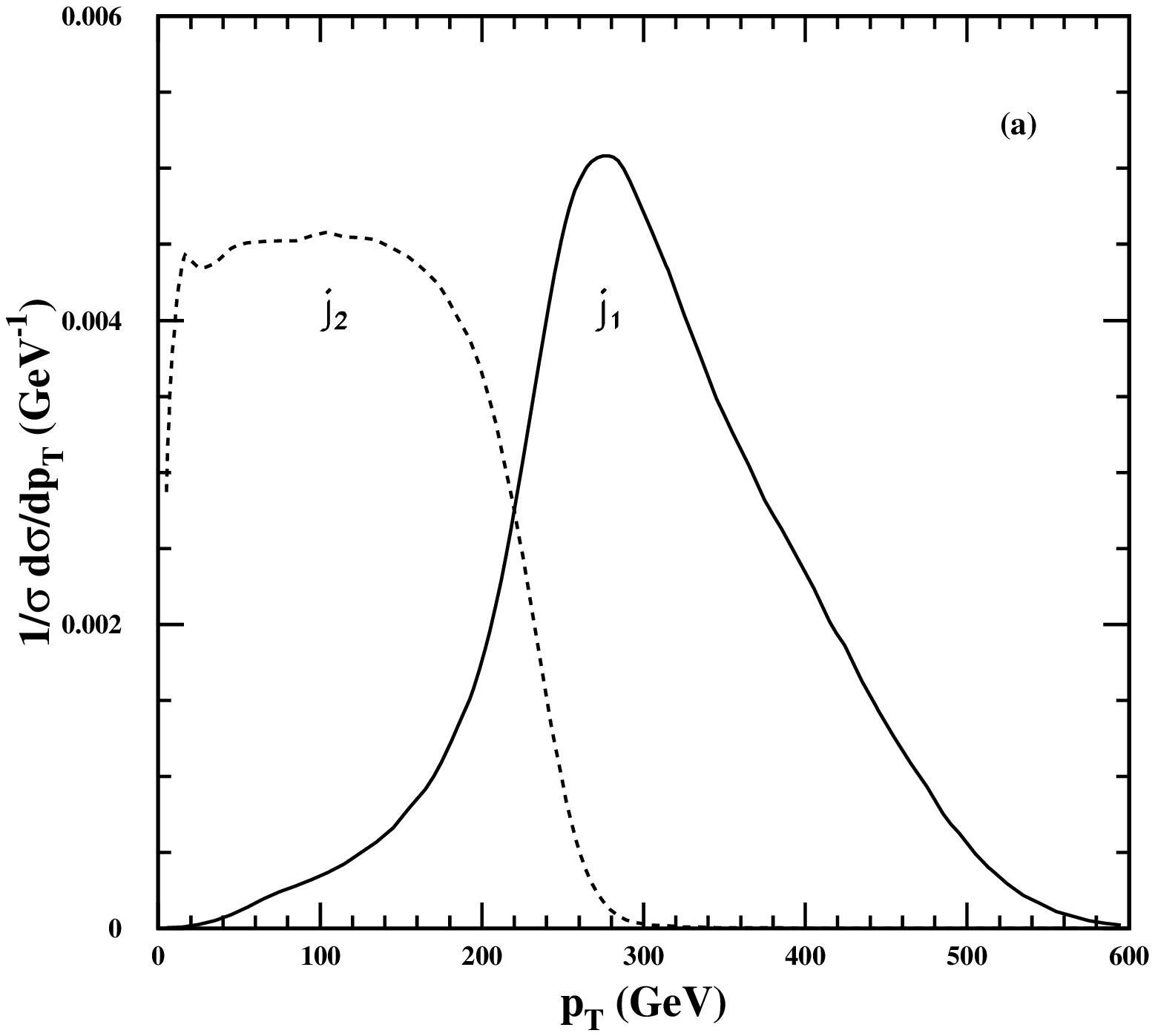}&
\includegraphics[clip,width=5cm,height=6cm,keepaspectratio]{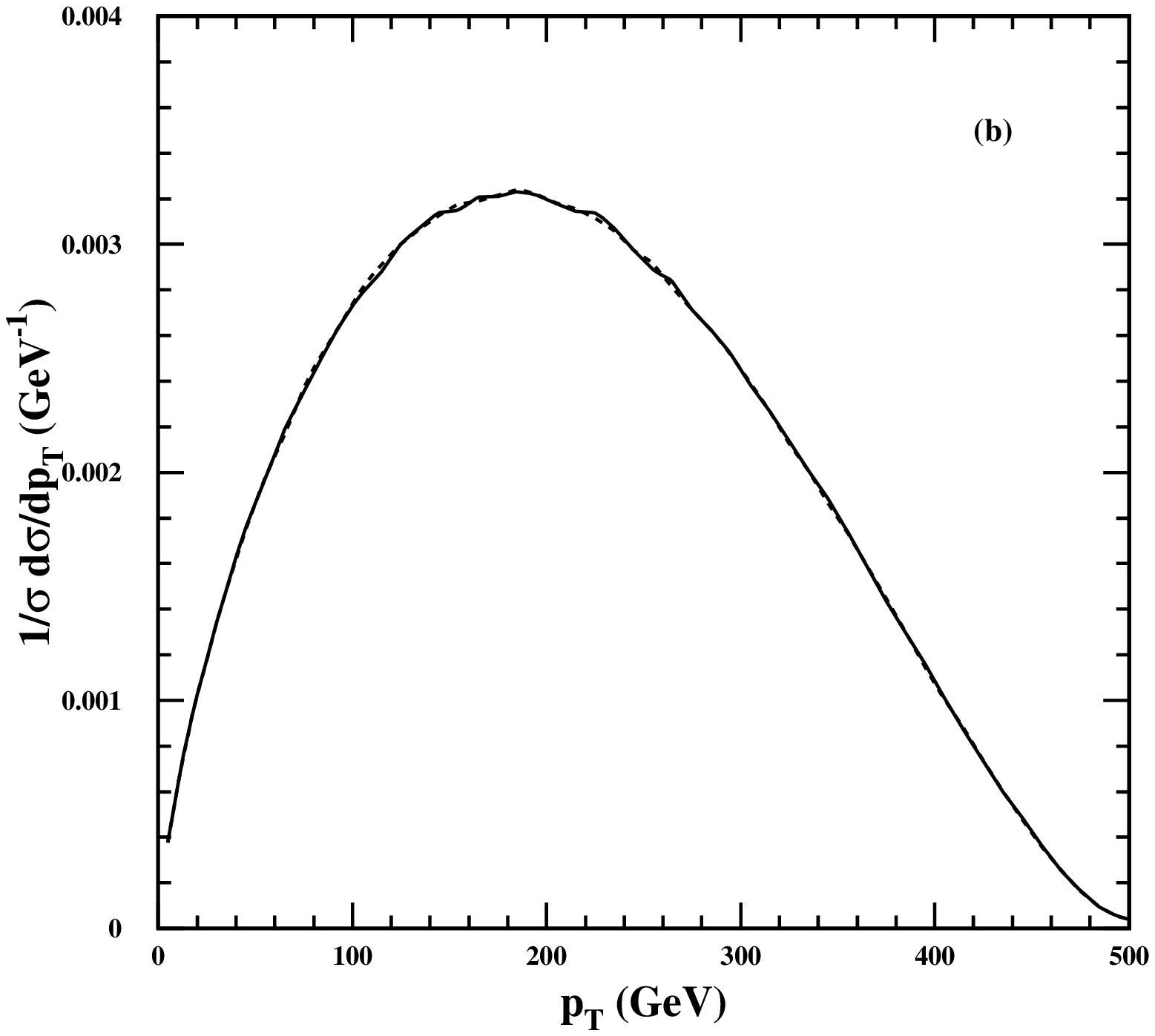}&
\includegraphics[clip,width=5cm,height=6cm,keepaspectratio]{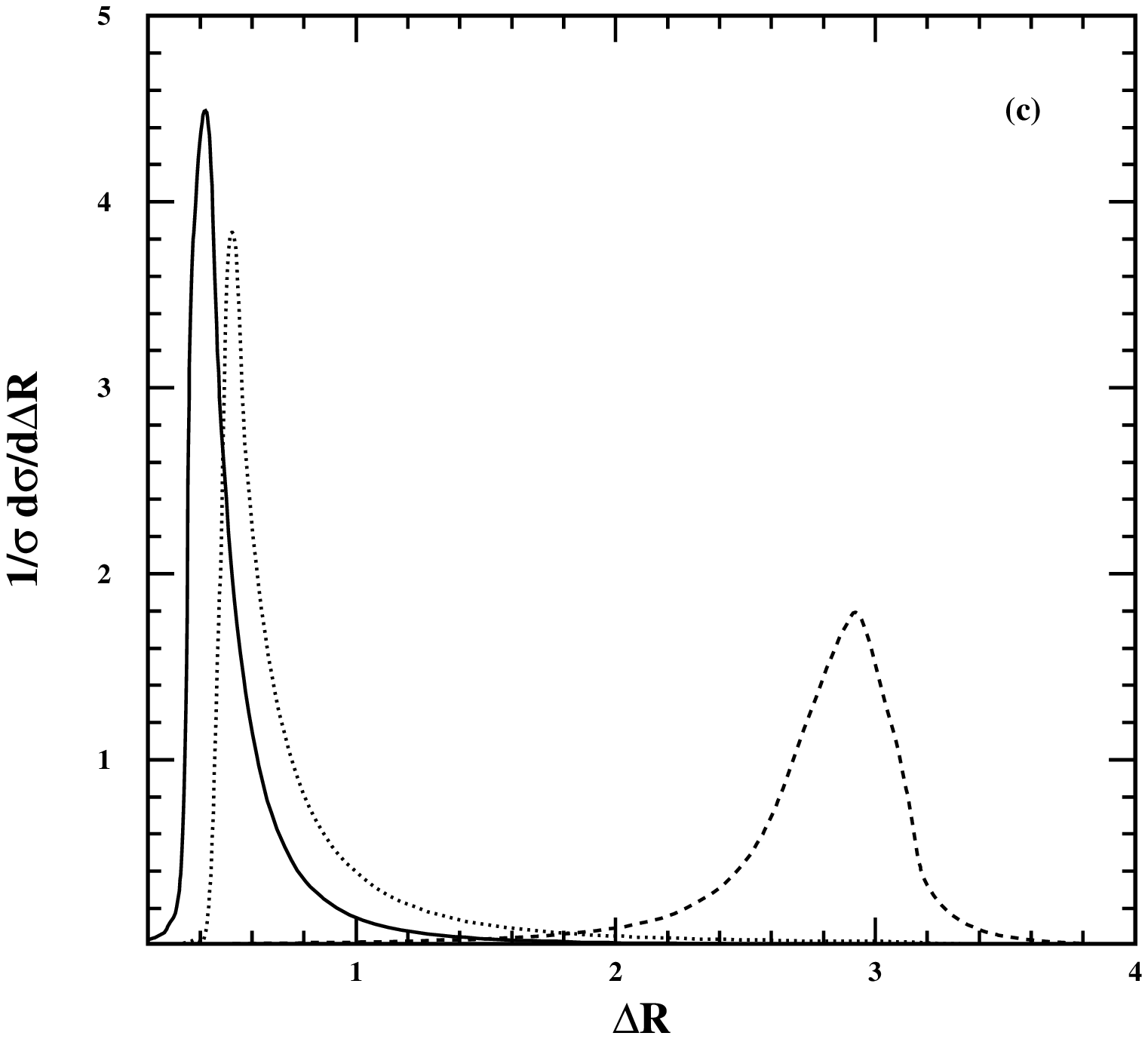}
\end{tabular}
\caption{(a) The transverse momentum distributions of the jets ($j_1,j_2$) with $p_{Tj_1}>p_{Tj_2}$ at $\sqrt{s}=14$ $TeV$. (b) The transverse momentum distributions of the leptons. (c) The minimal angular separation distributions between leptons (solid line), ~jets (dashed line) and that between jets and leptons (dotted line).}
\label{fig:pt}
\end{center}
\end{figure}
The transverse momentum distributions of the jets and leptons are shown in Fig.~\ref{fig:pt} (a) and (b). In order to identify the isolated jet (lepton), we define the angular separation between particle $i$
and particle $j$ as 
\begin{equation}
\Delta R_{ij}=\sqrt{\Delta\phi_{ij}^{2}+\Delta\eta_{ij}^{2}},
\end{equation}
 where $\Delta\phi_{ij}=\phi_{i}-\phi_{j}$ and $\Delta\eta_{ij}=\eta_{i}-\eta_{j}$.
$\phi_{i}$ ($\eta_{i}$) denotes the azimuthal angle (rapidity) of
the related jet or lepton. 

To be more realistic, the simulation at the detector is performed
by smearing the leptons and jets energies according to the assumption
of the Gaussian resolution parametrization 
\begin{equation}
\frac{\delta(E)}{E}=\frac{a}{\sqrt{E}}\oplus b,
\end{equation}
where $\delta(E)/E$ is the energy resolution, $a$ is a sampling
term, $b$ is a constant term, and $\oplus$ denotes a sum in quadrature.
We take $a=5\%$, $b=0.55\%$ for leptons and $a=100\%$, $b=5\%$
for jets respectively~\cite{Aad:2009wy,Ball:2007zza}.

The corresponding distributions for $\Delta R=min(\Delta R_{ij})$ are shown
in Fig.~\ref{fig:pt}~(c). Due to the large mass splitting between $Z'$ and Higgs as well as $Z$ boson, the mediated particles will be highly boosted and the jets (leptons) are mostly moving in the same direction with Higgs ($Z$), meanwhile jets are mostly moving in the opposite direction to leptons. According to the above distributions, we adopt the basic cuts (cut I)  
\begin{eqnarray}
&p_{T_l}>30~\it{GeV},~p_{T_{j1}}>100~\it{GeV},~p_{T_{j2}}>50~\it{GeV}, &\nonumber \\
&|\eta_l|<2.5,~|\eta_j|<3.0,~\Delta R_{jl}>2.0.&
\end{eqnarray}

\begin{figure}
\begin{center}
\begin{tabular}{ccc}
\includegraphics[clip,width=5cm,height=6cm,keepaspectratio]{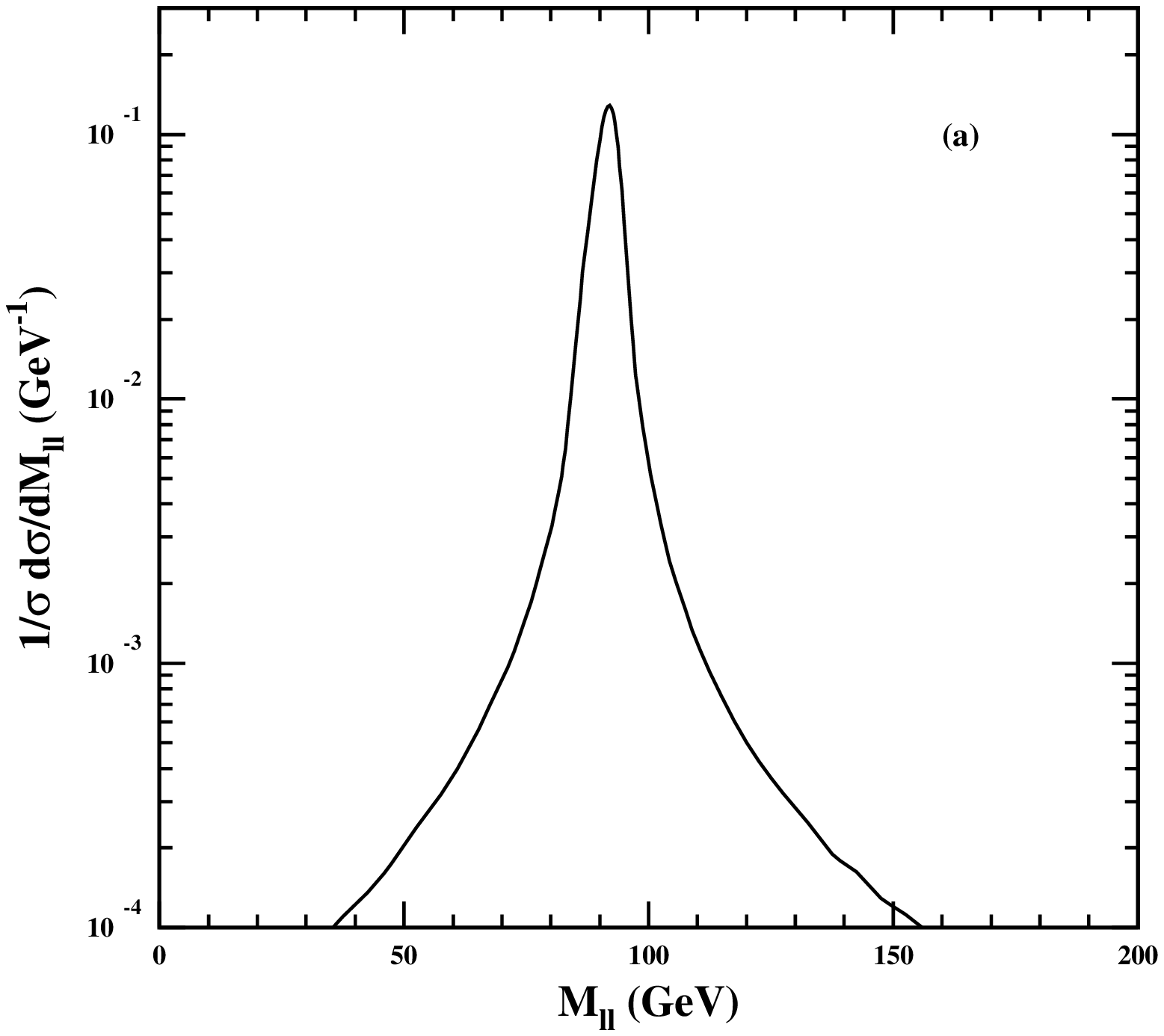}&
\includegraphics[clip,width=5cm,height=6cm,keepaspectratio]{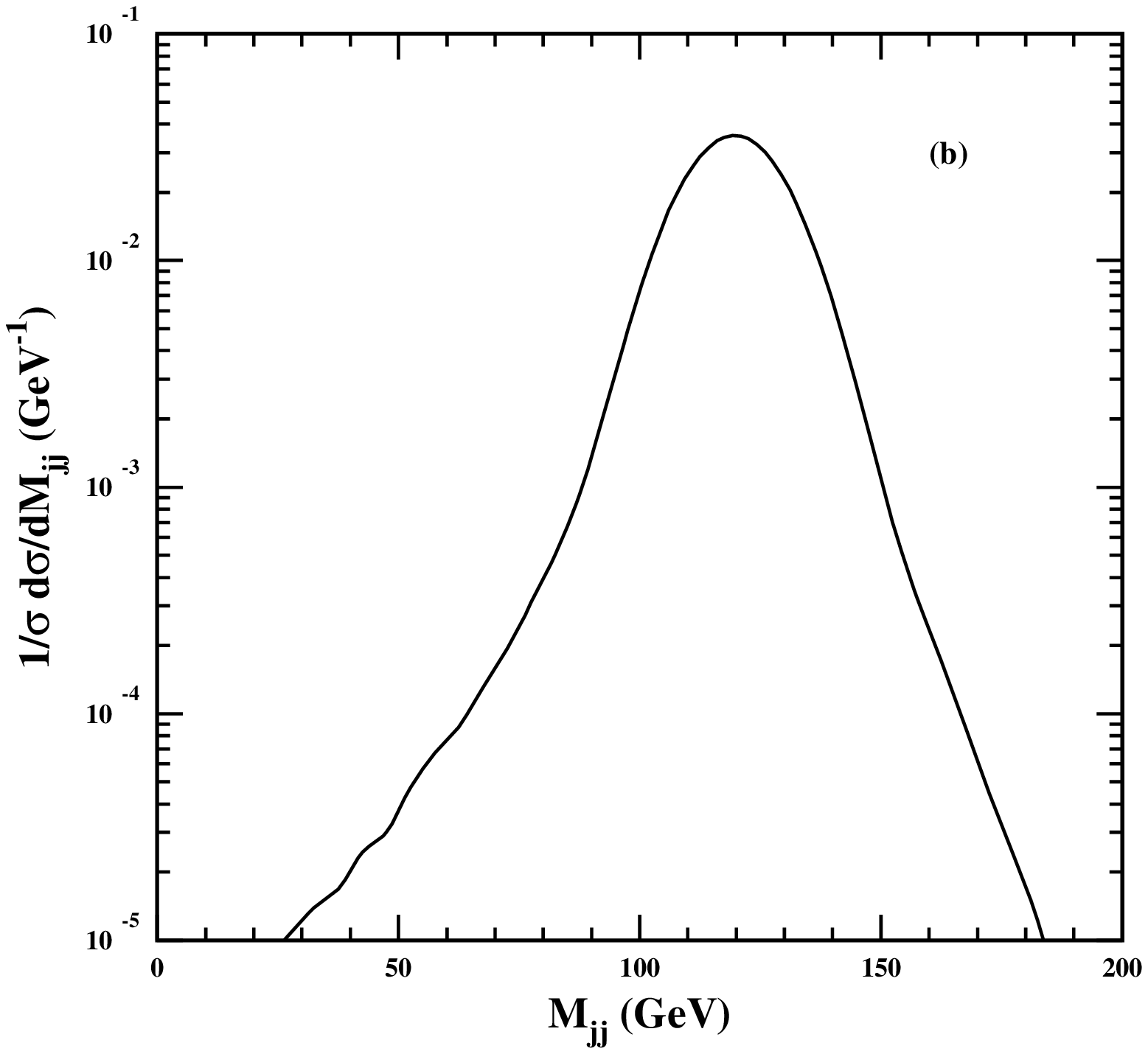}&
\includegraphics[clip,width=5cm,height=6cm,keepaspectratio]{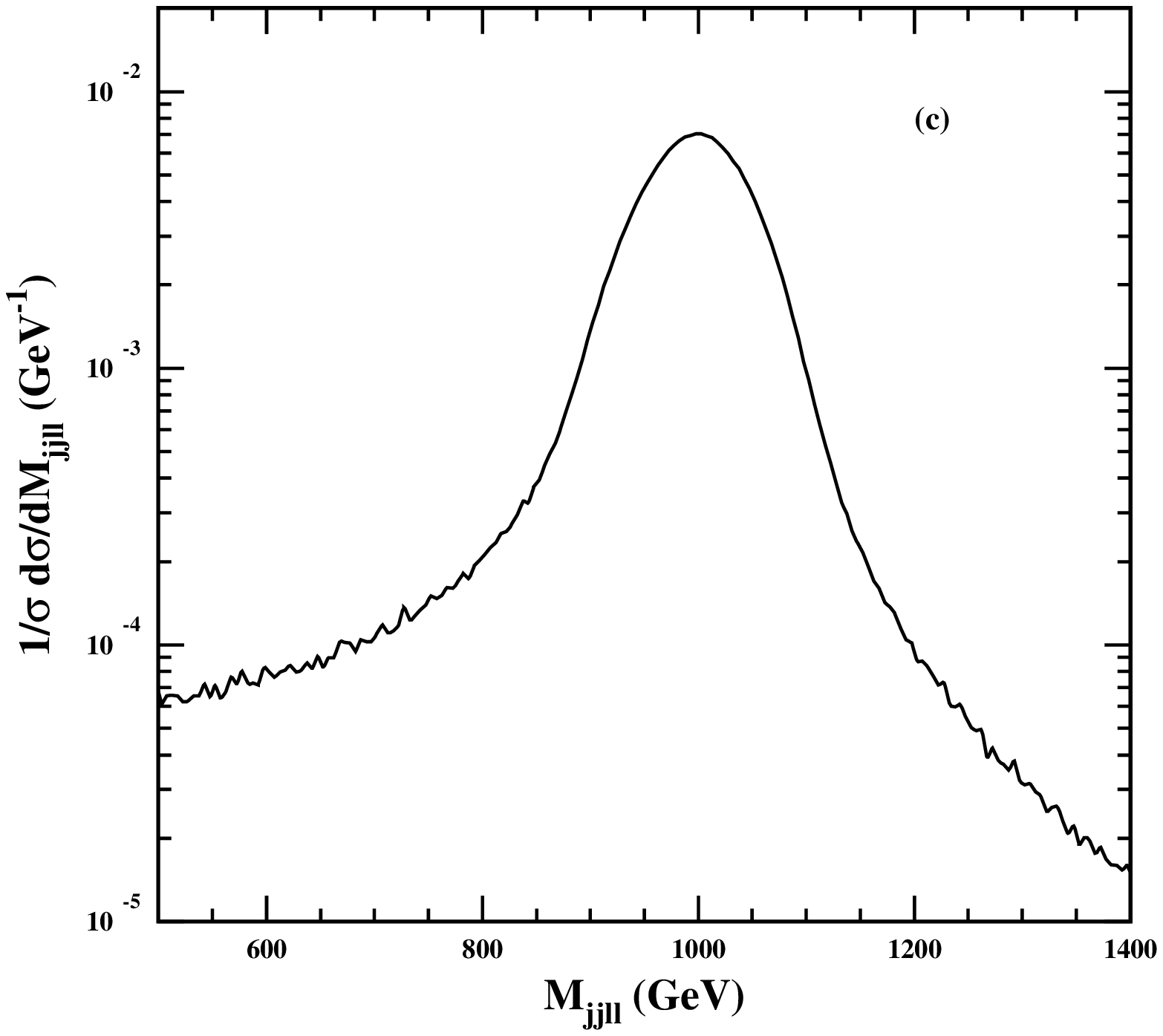}\\
(a)&(b)&(c)
\end{tabular}
\caption{The distribution for the invariant mass reconstructed from (a) leptons;  (b) jets; (c)  all final state particles.}
\label{fig:mass}
\end{center}
\end{figure}

Analyzing the $l^-l^+jj$ final state, we find that two leptons are from $Z$ boson and two jets are from Higgs boson. Hence, the mediate resonances can be reconstructed through the invariant masses of the jets and leptons respectively.  The invariant mass distributions reconstructed by (a) leptons, (b) jets and (c) leptons and jets are shown in Fig.~\ref{fig:mass} . We employ the invariant mass through various combinations of the final particles to constrain the mediate resonances (cut II),
 \begin{eqnarray}
\frac {|M_{jj}-m_H|}{m_H}\leq 10\%,~~\frac {|M_{ll}-m_Z|}{m_Z}\leq 10\%,	~ and ~\frac {|M_{jjll}-m_{Z'}|}{m_{Z'}}\leq 10\%,
\end{eqnarray}
together with one of the final jets tagged as b-jet.

Corresponding to the final state of $l^-l^+jj$, SM processes mediated by $~ZH,~t\bar t,~ZW,~ZZ,~Zjj,$ and $WWjj$ intermediate are the main backgrounds. To suppress the influence from the background processes, the discrepancies between signal and background processes are analyzed. Obviously, most of the cross section of $ZH$  is due to $Z$ boson resonance which can be cut down by adopting the final system invariant mass cut. The decay mode $t \bar t \to W^+ b W^- \bar b\to l^+ \bar \nu b l^- \nu b$ is required to obtain the $l^-l^+jj$ final state for $t \bar t$ process. It shows that the charged leptons are coming from different intermediate $W$ bosons in $t \bar t$ process, thus we can set the invariant mass reconstructed by two charged leptons to be around the $Z$ boson mass to cut down the $t \bar t$ process as well as $WWjj$ process. For the processes with jets, such as $Zjj$ and $WWjj$, jets are tremendously coming from QCD processes, then the invariant mass reconstructed by the final jets lead to different distributions compared with signal process. Besides, we require the final jets reconstructed invariant mass be around Higgs mass, $ZW$ and $ZZ$ process can be suppressed for the mass gaps between Higgs and $Z$ boson.
        
Furthermore, due to the $Z'$ mass hierarchy from Higgs and $Z$ boson, $Z$ boson reconstructed from the charged leptons will be with a high lorentz boost factor ($\gamma$). Thus we adopt
\begin{equation}
\gamma  = \frac{1}{\sqrt{1-v^2}} \ge 3.0,
\end{equation} 
referred as cut III to further suppress the backgrounds, where $v$ is the velocity of $Z$ boson in the laboratory frame. 

\begin{table}
\begin{tabular}{|c|c|c|c|c|} \hline
process&no cut& cut I & cut I+II & cut I+II+III\\ \hline
signal  &13.3&4.48&2.78&2.08\\ \hline
$ZH({\rm SM})$          &669 &0.30&0.04&0.03\\ \hline
$t\bar t$& $5.38 \times 10^{5}$  &20.5&0.39&-\\ \hline
$ZW$          &$2.76 \times 10^{4}$ &2.68&-&-\\ \hline
$ZZ$          &$1.05 \times 10^{4}$ &3.13&-&-\\ \hline
$Zjj$ &$7.37 \times 10^{6}$ &855 &-&-\\ \hline
$W Wjj$ &$6.88 \times 10^{4}$ &10.7 &-&-\\ \hline
\end{tabular}
\caption{The total cross section ($fb$) for signal and backgrounds process before and after cuts  at the $\sqrt{s} = 14$ TeV  LHC with $m_{Z'}=1$ TeV.}
\label{Tab:background}
\end{table}
\begin{table}
\begin{tabular}{|c|c|c|c|c|} \hline
$m_{Z'}$&no cut& cut I & cut I+II & cut I+II+III\\ \hline
 1~TeV  &13.3&4.48&2.78&2.08\\ \hline
1.5~TeV  &9.67&1.14&0.56&0.42\\ \hline
2~TeV  &9.31&0.60&0.15&0.11\\ \hline
\end{tabular}
\caption{The total cross section ($fb$) for signal process before and after cuts  at the $\sqrt{s} = 14$ TeV  LHC with $m_{Z'}=1, ~1.5, ~2$ TeV.}
\label{Tab:cs}
\end{table}
\begin{figure}
\begin{center}
\begin{tabular}{c}
\includegraphics[clip,width=14cm,height=10cm,keepaspectratio]{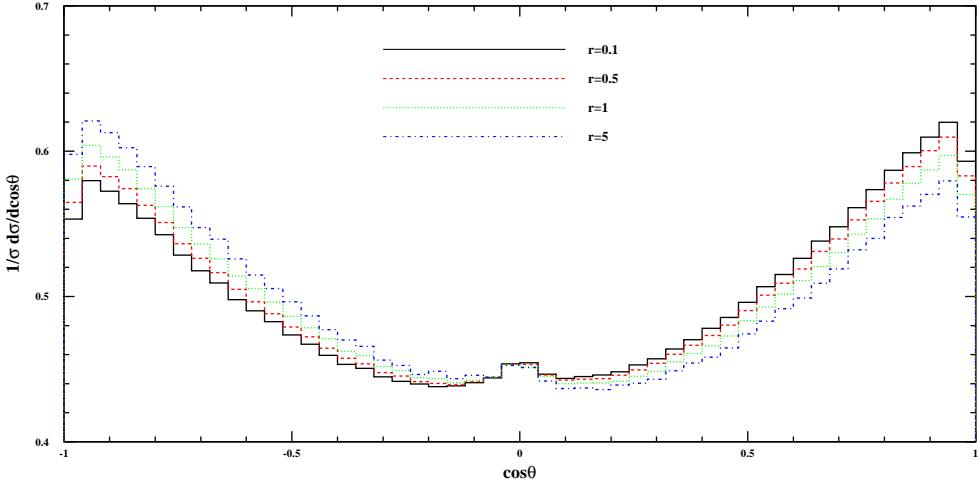}
\end{tabular}
\caption{The angular distribution for charged lepton after all cuts at LHC with $m_{Z'}=1.5$ TeV.}
\label{fig:theta-r2}
\end{center}
\end{figure}

The total cross sections for the signal and background processes before and after cuts are listed in Table~\ref{Tab:background}. The dominant backgrounds $t\bar t$ and $Zjj$  cross section are over $\cal{O}$($10^4$)  than the signal process before cuts. After the basic kinematic cuts and the invariant mass cuts (cut I$+$II), most backgrounds are significantly suppressed. $t\bar t$ process will be further suppressed by cut III. 
With SM ZH process substantially reduced, all the backgrounds are eliminated through the three cuts, thus the signal process can be clearly studied. The cross sections for the other choice of $Z'$ mass are also listed in Table~\ref{Tab:cs}. It shows that more than one hundred events can be detected at LHC with a luminosity of 300 $fb^{-1}$ for $m_{Z'}=1.5$ TeV. The background processes can be much strongly suppressed with larger final system invariant mass and  a loose $M_{jjll}$ cut can be usd to emphasize the signal.

After adopting all the kinematic cuts, we display the $1/\sigma d\sigma/d\cos \theta$ distributions  versus $\cos \theta$ for different $r$ in Fig.～\ref{fig:theta-r2}. The distributions with $|\cos \theta|$ close to zero and one are distorted by the kinematic cuts to some extent, but one can find that the charged leptons tend to move along with the opposite direction of the final system moving direction with large $r$.  It is possible to utilize the angular distribution to distinguish various models including $Z'q\bar q$ interaction with different chiral couplings.  
   
Corresponding to the charged lepton angular distribution of process~(\ref{eq:pp2llbb}), we define a kind of forward-backward asymmetry ($A_{FB}$),
\begin{equation}
A_{FB}=\frac {\sigma(cos\theta \geq 0)-\sigma(cos\theta<0)}{\sigma(cos\theta \geq 0)+\sigma(cos\theta<0)}.
\end{equation}
The cross sections and $A_{FB}$ are listed in Table \ref{Tab:A_FB}. It is obvious that the forward-backward asymmetry is sensitive to different $r$. With the increasing of $r$ value,  $A_{FB}$ changes from negative to positive. The absolute value of forward-backward asymmetry becomes smaller for heavier $Z'$ with fixed $r$. The numbers in Table \ref{Tab:A_FB} show that with an integrated luminosity of 300 $fb^{-1}$ for center-of-mass at 14 TeV, 670 (130) events can be expected  for $Z'$ mass at 1 (1.5) TeV with $r=1$, and $A_{FB}$ can reach 0.032 (-0.032) for $r=0.1$ ($r=5$) with $m_{Z'}=1$ TeV. One can find that after the acceptance cuts, the angular distribution  with respect to $\cos\theta$ and $A_{FB}$ is helpful to investigate the $Z'ZH$ interaction via $ZH$ associated production at LHC. 

\begin{table}
\begin{tabular}{|c|c|c|c|c|c|c|c|c|c|c|c|c|} \hline
$m_{Z'}$(TeV)&\multicolumn{4}{c|} {1}  &\multicolumn{4}{c|} {1.5} &\multicolumn{4}{c|} {2} \\ \hline
r&r=0.1&r=0.5&r=1&r=5&r=0.1&r=0.5&r=1&r=5&r=0.1&r=0.5&r=1&r=5\\ \hline
$\sigma$($fb$)&2.10&2.20&2.25&0.54&0.42&0.44&0.45&0.11&0.11&0.11&0.12&0.03\\ \hline
$A_{FB}$&0.032&0.016&-0.006&-0.032&0.022&0.011&-0.004&-0.021&0.016&0.008&-0.003&-0.016\\ \hline
\end{tabular}
\caption{The total cross section ($\sigma$) and forward-backward asymmetry ($A_{FB}$) after cuts at the LHC with $\sqrt{s} = 14$ TeV.}
\label{Tab:A_FB}
\end{table}
%
\section{Summary}\label{summary}
Many extensions beyond SM predict the existence of new heavy neutral gauge boson.  The recently discovered SM-like Higgs boson at LHC motivates us to investigate the $Z'ZH$ interaction.  We study the process of $pp\rightarrow Z'\rightarrow ZH $ with $Z\rightarrow l^+l^-$ and  $H\rightarrow b\bar b$ decay modes. A massive $Z'$ boson can be reconstructed through the resonance peak that appears in the invariant mass spectrum of the final states $l^+l^-b\bar b$.  The couplings of $Z'$ to SM fermions can be extracted from right-handed and left-handed current, thus the chiral coupling ratio $r$ can serve as an important parameter to specify various models. We find that the angular distribution for the final lepton can be related to $r$ via the $Z'ZH$ interaction. The backgrounds from SM with the same final state are estimated and efficiently suppressed by the kinematic cuts. Corresponding to the angular distribution of final state charged lepton, we define a forward-backward asymmetry $A_{FB}$ which is 
sensitive to $r$. A variety of new physics models predict different chiral coupling ratio $r$, thus the charged lepton angular distribution and the forward-backward asymmetry can help to understand the $Z'ZH$ interaction and distinguish between different models.

\begin{acknowledgments}
This work is supported in part by the NSFC and NSC. The authors would like to thank Profs.  X. He,  S. Li, S. Bao and Y. Jin for their useful comments and helpful discussions on this work.
\end{acknowledgments}

\end{document}